\documentclass[aip,graphicx]{revtex4-1}

\draft % marks overfull lines with a black rule on the right

\begin{document}

\title{The Monopole and the Coulomb field as duals within the unifying Reissner-Nordstr\"{o}m geometry} %Title of paper

\author{Alcides Garat}
%\thanks{}
\affiliation{1. Instituto de F\'{\i}sica, Facultad de Ciencias,
Igu\'a 4225, esq. Mataojo, Montevideo, Uruguay.}

\date{April 18th, 2012}

\begin{abstract}
We are going to prove that the Monopole and the Coulomb fields are duals within the unifying structure provided by the Reissner-Nordstr\"{o}m spacetime. This is accomplished when noticing that in order to produce the tetrad that locally and covariantly diagonalizes the stress-energy tensor, both the Monopole and the Coulomb fields are necessary in the construction. Without any of them it would be impossible to express the tetrad vectors that locally and covariantly diagonalize the stress-energy tensor. Then, both electromagnetic fields are an integral part of the same structure, the Reissner-Nordstr\"{o}m geometry.
\end{abstract}


\maketitle %\maketitle must follow title, authors, abstract and \pacs

\section{Introduction}

For this purpose of proving that both electromagnetic fields are an integral part of the same structure, the Reissner-Nordstr\"{o}m geometry, we have to review the results found in a previous work like manuscript\cite{A}. In this paper we found that locally the electromagnetic gauge group of transformations was isomorphic to tetrad Lorentz transformations in both orthogonal planes or blades, one and two \cite{SCH}. That is to say, isomorphic to local Lorentz transformations on both planes, independently. It was found that in a curved four dimensional Lorentzian spacetime where a non-null electromagnetic field is present, a tetrad can be built such that these vectors covariantly diagonalize the stress-energy tensor at every point in spacetime. Therefore, the symmetry represented by local electromagnetic gauge transformations can be thought of as Lorentz transformations of the tetrad unit vectors inside these blades. Blade one is generated by a timelike and a spacelike vectors. Blade two by the other two spacelike vectors. It is in the context of all the results found in reference \cite{A} that we are going to argue the following. When we build these tetrad vectors we need in the particular case of the Reissner-Nordstr\"{o}m spacetime both dual solutions, the Monopole and the Coulomb field. Without any of them this construction would be impossible. Since both planes spanned by the pairs of tetrad vectors that locally and covariantly diagonalize the electromagnetic stress-energy tensor are unique, we have to argue that the Monopole and the Coulomb field are components of a unified structure. In section \ref{reissnordtetrad} we will introduce the subject of local symmetry in terms of Lorentz tetrad transformations on blades one and two through the Reissner-Nordstr\"{o}m geometry. The necessity for both electromagnetic fields within the unifying Reissner-Nordstr\"{o}m spacetime will arise in a natural way. Throughout the paper we use the conventions of paper \cite{MW}. In particular we use a metric with sign conventions -+++. The only difference in notation with \cite{MW} will be that we will call our geometrized electromagnetic potential $A^{\alpha}$, where $f_{\mu\nu}=A_{\nu ;\mu} - A_{\mu ;\nu}$ is the geometrized electromagnetic field $f_{\mu\nu}= (G^{1/2} / c^2) \: F_{\mu\nu}$.

\section{The Reissner-Nordstr\"{o}m geometry and the new tetrad}
\label{reissnordtetrad}

The line element for this spacetime is given by the following expression \cite{RW}$^{,}$\cite{MC2},

\begin{eqnarray}
ds^{2} = - (1 - {2m \over r} + {q^{2} \over r^{2}})\: dt^{2} + (1 - {2m \over r} + {q^{2} \over r^{2}})^{-1}\: dr^{2} + r^{2}\:(d\theta^{2} + \sin^{2}\theta\:d\phi^{2})\ . \label{reissnord}
\end{eqnarray}

We will also introduce the vectors that diagonalize locally and covariantly the stress-energy tensor and generate blades one and two. Tetrads become tools of primary importance, as local gauge symmetries are associated to structures that can be expressed in terms of these new tetrad vectors. We present first, the four tetrad vectors introduced in paper \cite{A} that locally and covariantly diagonalize the electromagnetic stress-energy tensor and define at every point in spacetime the blades one and two.

\begin{eqnarray}
V_{(1)}^{\alpha} &=& \xi^{\alpha\lambda}\:\xi_{\rho\lambda}\:X^{\rho}
\label{V1}\\
V_{(2)}^{\alpha} &=&  \sqrt{-Q/2} \:\: \xi^{\alpha\lambda} \: X_{\lambda}
\label{V2}\\
V_{(3)}^{\alpha} &=& \sqrt{-Q/2} \:\: \ast \xi^{\alpha\lambda} \: Y_{\lambda}
\label{V3}\\
V_{(4)}^{\alpha} &=& \ast \xi^{\alpha\lambda}\: \ast \xi_{\rho\lambda}
\:Y^{\rho}\ .\label{V4}
\end{eqnarray}

where $Q=\xi_{\mu\nu}\:\xi^{\mu\nu}=-\sqrt{T_{\mu\nu}T^{\mu\nu}}$ according to equations (39) in \cite{MW}. $Q$ is assumed not to be zero, because we are dealing with non-null electromagnetic fields. The first two (\ref{V1}-\ref{V2}) eigenvectors of the stress-energy tensor with eigenvalue $Q/2$, the last two (\ref{V3}-\ref{V4}) with eigenvalue $-Q/2$. We briefly remind ourselves that the original expression for the electromagnetic stress-energy tensor $T_{\mu\nu}= f_{\mu\lambda}\:\:f_{\nu}^{\:\:\:\lambda} + \ast f_{\mu\lambda}\:\ast f_{\nu}^{\:\:\:\lambda}$ is given in terms of the electromagnetic tensor $f_{\mu\nu}$ and its dual $\ast f_{\mu\nu}={1 \over 2}\:\epsilon_{\mu\nu\sigma\tau}\:f^{\sigma\tau}$. After a local duality transformation,

\begin{equation}
f_{\mu\nu} = \xi_{\mu\nu} \: \cos\alpha +
\ast\xi_{\mu\nu} \: \sin\alpha\ ,\label{dr}
\end{equation}

where the local scalar $\alpha$ is the complexion, we are able to write the stress-energy in terms of the extremal field $\xi_{\mu\nu}$ and its dual. We can express the extremal field as,

\begin{equation}
\xi_{\mu\nu} = e^{-\ast \alpha} f_{\mu\nu}\ = \cos\alpha\:f_{\mu\nu} - \sin\alpha\:\ast f_{\mu\nu}.\label{dref}
\end{equation}

Extremal fields are essentially electric fields and they satisfy,

\begin{equation}
\xi_{\mu\nu} \ast \xi^{\mu\nu}= 0\ . \label{i0}
\end{equation}

Equation (\ref{i0}) is a condition imposed on (\ref{dref}) and then the explicit expression for the complexion emerges $\tan(2\alpha) = - f_{\mu\nu}\:\ast f^{\mu\nu} / f_{\lambda\rho}\:f^{\lambda\rho}$. As antisymmetric fields in a four dimensional Lorentzian spacetime, the extremal fields also verify the identity,

\begin{eqnarray}
\xi_{\mu\alpha}\:\xi^{\nu\alpha} -
\ast \xi_{\mu\alpha}\: \ast \xi^{\nu\alpha} &=& \frac{1}{2}
\: \delta_{\mu}^{\:\:\nu}\ Q \ ,\label{i1}
\end{eqnarray}

It can be proved that condition (\ref{i0}) and through the use of the general identity,

\begin{eqnarray}
A_{\mu\alpha}\:B^{\nu\alpha} -
\ast B_{\mu\alpha}\: \ast A^{\nu\alpha} &=& \frac{1}{2}
\: \delta_{\mu}^{\:\:\nu}\: A_{\alpha\beta}\:B^{\alpha\beta}  \ ,\label{ig}
\end{eqnarray}

which is valid for every pair of antisymmetric tensors in a four-dimensional Lorentzian spacetime \cite{MW}, when applied to the case $A_{\mu\alpha} = \xi_{\mu\alpha}$ and $B^{\nu\alpha} = \ast \xi^{\nu\alpha}$ yields the equivalent condition,

\begin{eqnarray}
\xi_{\alpha\mu}\:\ast \xi^{\mu\nu} &=& 0\ ,\label{i2}
\end{eqnarray}

which is equation (64) in \cite{MW}. It is evident that identity (\ref{i1}) is a special case of (\ref{ig}). The duality rotation given by equation (\ref{dr}) allows us to express the stress-energy tensor in terms of the extremal field,

\begin{equation}
T_{\mu\nu}=\xi_{\mu\lambda}\:\:\xi_{\nu}^{\:\:\:\lambda}
+ \ast \xi_{\mu\lambda}\:\ast \xi_{\nu}^{\:\:\:\lambda}\ .\label{TEMDR}
\end{equation}

With all these elements it becomes trivial to prove that the tetrad (\ref{V1}-\ref{V4}) is orthogonal and diagonalizes the stress-energy tensor (\ref{TEMDR}). We notice then that we still have to define the vectors $X^{\mu}$ and $Y^{\mu}$. Let us introduce some names. The tetrad vectors have two essential components. For instance in vector $V_{(1)}^{\alpha}$ there are two main structures. First, the skeleton, in this case $\xi^{\alpha\lambda}\:\xi_{\rho\lambda}$, and second, the gauge vector $X^{\rho}$. The gauge vectors it was proved in manuscript \cite{A} could be anything that does not make the tetrad vectors trivial. That is, the tetrad (\ref{V1}-\ref{V4}) diagonalizes the stress-energy tensor for any non-trivial gauge vectors $X^{\mu}$ and $Y^{\mu}$. It was therefore proved that we can make different choices for $X^{\mu}$ and $Y^{\mu}$. In geometrodynamics, the Maxwell equations,

\begin{eqnarray}
f^{\mu\nu}_{\:\:\:\:\:;\nu} &=& 0 \label{L1}\nonumber\\
\ast f^{\mu\nu}_{\:\:\:\:\:;\nu} &=& 0 \ , \label{L2}
\end{eqnarray}

are telling us that two potential vector fields $A_{\nu}$ and $\ast A_{\nu}$ exist,

\begin{eqnarray}
f_{\mu\nu} &=& A_{\nu ;\mu} - A_{\mu ;\nu}\label{ER}\nonumber\\
\ast f_{\mu\nu} &=& \ast A_{\nu ;\mu} - \ast A_{\mu ;\nu} \ .\label{DER}
\end{eqnarray}

The symbol $``;''$ stands for covariant derivative with respect to the metric tensor $g_{\mu\nu}$. We can define then, a normalized tetrad with the choice $X^{\mu} = A^{\mu}$ and $Y^{\mu} = \ast A^{\mu}$,

\begin{eqnarray}
U^{\alpha} &=& \xi^{\alpha\lambda}\:\xi_{\rho\lambda}\:A^{\rho} \:
/ \: (\: \sqrt{-Q/2} \: \sqrt{A_{\mu} \ \xi^{\mu\sigma} \
\xi_{\nu\sigma} \ A^{\nu}}\:) \label{U}\\
V^{\alpha} &=& \xi^{\alpha\lambda}\:A_{\lambda} \:
/ \: (\:\sqrt{A_{\mu} \ \xi^{\mu\sigma} \
\xi_{\nu\sigma} \ A^{\nu}}\:) \label{V}\\
Z^{\alpha} &=& \ast \xi^{\alpha\lambda} \: \ast A_{\lambda} \:
/ \: (\:\sqrt{\ast A_{\mu}  \ast \xi^{\mu\sigma}
\ast \xi_{\nu\sigma}  \ast A^{\nu}}\:)
\label{Z}\\
W^{\alpha} &=& \ast \xi^{\alpha\lambda}\: \ast \xi_{\rho\lambda}
\:\ast A^{\rho} \: / \: (\:\sqrt{-Q/2} \: \sqrt{\ast A_{\mu}
\ast \xi^{\mu\sigma} \ast \xi_{\nu\sigma} \ast A^{\nu}}\:) \ .
\label{W}
\end{eqnarray}

The four vectors (\ref{U}-\ref{W}) have the following algebraic properties,

\begin{equation}
-U^{\alpha}\:U_{\alpha}=V^{\alpha}\:V_{\alpha}
=Z^{\alpha}\:Z_{\alpha}=W^{\alpha}\:W_{\alpha}=1 \ .\label{TSPAUX}
\end{equation}

Using the equations (\ref{i1}-\ref{i2}) it is simple to prove that (\ref{U}-\ref{W}) are orthonormal. When we make the transformation,

\begin{eqnarray}
A_{\alpha} \rightarrow A_{\alpha} + \Lambda_{,\alpha}\ , \label{G1}
\end{eqnarray}

$f_{\mu\nu}$ remains invariant, and the transformation,

\begin{eqnarray}
\ast A_{\alpha} \rightarrow \ast A_{\alpha} +
\ast \Lambda_{,\alpha}\ , \label{G2}
\end{eqnarray}

leaves $\ast f_{\mu\nu}$ invariant, as long as the functions $\Lambda$ and $\ast \Lambda$ are scalars. Schouten \cite{SCH} defined what he called, a two-bladed structure in a spacetime \cite{SCH}. These blades are the planes determined by the pairs ($U^{\alpha}, V^{\alpha}$) and ($Z^{\alpha}, W^{\alpha}$).
It was proved in \cite{A} that the transformation (\ref{G1}) generates a ``rotation'' of the tetrad vectors ($U^{\alpha}, V^{\alpha}$) into ($\tilde{U}^{\alpha}, \tilde{V}^{\alpha}$) such that these ``rotated'' vectors ($\tilde{U}^{\alpha}, \tilde{V}^{\alpha}$) remain in the plane or blade one generated by ($U^{\alpha}, V^{\alpha}$). It was also proved in \cite{A} that the transformation (\ref{G2}) generates a ``rotation'' of the tetrad vectors ($Z^{\alpha}, W^{\alpha}$) into ($\tilde{Z}^{\alpha}, \tilde{W}^{\alpha}$) such that these ``rotated'' vectors ($\tilde{Z}^{\alpha}, \tilde{W}^{\alpha}$) remain in the plane or blade two generated by ($Z^{\alpha}, W^{\alpha}$).  For example, a boost of the two vectors $(U^{\alpha},\:V^{\alpha})$ on blade one, given in (\ref{U}-\ref{V}), by the ``angle'' $\phi$ can be written,

\begin{eqnarray}
U^{\alpha}_{(\phi)}  &=& \cosh(\phi)\: U^{\alpha} +  \sinh(\phi)\: V^{\alpha} \label{UT} \\
V^{\alpha}_{(\phi)} &=& \sinh(\phi)\: U^{\alpha} +  \cosh(\phi)\: V^{\alpha} \label{VT} \ .
\end{eqnarray}

There are also discrete transformations of vectors $(U^{\alpha},\:V^{\alpha})$ on blade one, see reference \cite{A}. The rotation of the two tetrad vectors $(Z^{\alpha},\:W^{\alpha})$ on blade two, given in (\ref{Z}-\ref{W}), by the ``angle'' $\varphi$, can be expressed as,

\begin{eqnarray}
Z^{\alpha}_{(\varphi)}  &=& \cos(\varphi)\: Z^{\alpha} -  \sin(\varphi)\: W^{\alpha} \label{ZT} \\
W^{\alpha}_{(\varphi)}  &=& \sin(\varphi)\: Z^{\alpha} +  \cos(\varphi)\: W^{\alpha} \label{WT} \ .
\end{eqnarray}

It is a simple exercise in algebra to see that the equalities $U^{[\alpha}_{(\phi)}\:V^{\beta]}_{(\phi)} = U^{[\alpha}\:V^{\beta]}$ and $Z^{[\alpha}_{(\varphi)}\:W^{\beta]}_{(\varphi)} = Z^{[\alpha}\:W^{\beta]}$ are true. These equalities are telling us that these antisymmetric tetrad objects are gauge invariant. We remind ourselves that it was proved in manuscript \cite{A} that the group of local electromagnetic gauge transformations is isomorphic to the local group LB1 of boosts plus discrete transformations on blade one, and independently to LB2, the local group of spatial rotations on blade two. Equations (\ref{UT}-\ref{VT}) represent a local electromagnetic gauge transformation of the vectors $(U^{\alpha}, V^{\alpha})$. Equations (\ref{ZT}-\ref{WT}) represent a local electromagnetic gauge transformation of the vectors $(Z^{\alpha}, W^{\alpha})$. Written in terms of these tetrad vectors, the electromagnetic field is,

\begin{equation}
f_{\alpha\beta} = -2\:\sqrt{-Q/2}\:\:\cos\alpha\:\:U_{[\alpha}\:V_{\beta]} +
2\:\sqrt{-Q/2}\:\:\sin\alpha\:\:Z_{[\alpha}\:W_{\beta]}\ .\label{EMF}
\end{equation}

Equation (\ref{EMF}) represents maximum simplification in the expression of the electromagnetic field. The true degrees of freedom are the local scalars $\sqrt{-Q/2}$ and $\alpha$. Local gauge invariance is manifested explicitly through the possibility of ``rotating'' through a scalar angle $\phi$ on blade one by a local gauge transformation (\ref{UT}-\ref{VT}) the tetrad vectors $U^{\alpha}$ and $V^{\alpha}$, such that
$U_{[\alpha}\:V_{\beta]}$ remains invariant \cite{A}. Analogous for discrete transformations on blade one. Similar analysis on blade two. A spatial ``rotation'' of the tetrad vectors $Z^{\alpha}$ and $W^{\alpha}$ through an ``angle'' $\varphi$ as in (\ref{ZT}-\ref{WT}), such that $Z_{[\alpha}\:W_{\beta]}$ remains invariant \cite{A}. All this formalism clearly provides a technique to maximally simplify the expression for the electromagnetic field strength. We finally conclude in this brief preview, that by transitivity it was proven that the boosts plus discrete transformations on plane one are isomorphic to the spatial rotations on plane two. We proceed to apply all this geometrical elements to the Reissner-Nordstr\"{o}m case with the choice $X^{\rho} = A^{\rho}$ and $Y^{\rho} = \ast A^{\rho}$, where the symbol $\ast$ in this particular last case is not the Hodge operator but a name. In the standard spherical coordinates $t, r, \theta, \phi$ the only non-zero components for the potentials will be $A_{t} = - q / r$ and $\ast A_{\phi} = -q\:\cos\theta$. With these potentials we find that the only non-zero components for the electromagnetic tensor $f_{\mu\nu} = A_{\nu ;\mu} - A_{\mu ;\nu}$ and its Hodge dual $\ast f_{\mu\nu} = \ast A_{\nu ;\mu} - \ast A_{\mu ;\nu}$ are $f_{tr} = - q / r^{2}$ and $\ast f_{\theta\phi} = q\:\sin\theta$. The symbol $;$ stands for covariant derivative with respect to the metric tensor $g_{\mu\nu}$, in our case the Reissner-Nordstr\"{o}m geometry. It is easy to check that the only non-zero components of the extremal field and its dual are $\xi_{tr} = f_{tr}$ and $\ast\xi_{\theta\phi} = \ast f_{\theta\phi}$. This is due to the fact that for the Reissner-Nordstr\"{o}m geometry $\tan(2\alpha)=0$. Since the extremal field and its dual are gauge invariants, their expression is unique. Therefore, when we observe the skeletons in tetrad vectors (\ref{U}-\ref{W}), we notice that these skeletons are unique. Precisely because of their local gauge invariance, and also because they also diagonalize locally and covariantly  the stress-energy tensor in a unique fashion. Then, it is evident that we need both the Coulomb and the Monopole field in order to implement their construction. Both electromagnetic fields simultaneously. We proceed again to write explicitly the only non-zero components of vectors (\ref{U}-\ref{W}),

\begin{eqnarray}
U^{t} &=& - (\sqrt{q^{2}}/q) / \sqrt{1 - {2m \over r} + {q^{2} \over r^{2}}} \label{Ut}\\
V^{r} &=& \sqrt{1 - {2m \over r} + {q^{2} \over r^{2}}} \label{Vr}\\
Z^{\theta} &=& - \sqrt{\cos^{2}\theta} / (r\:\cos\theta)  \label{Ztheta}\\
W^{\phi} &=& - \sqrt{q^{2}}\:\sqrt{\cos^{2}\theta} / (q\:r\:\sin\theta\:\cos\theta) \ .\label{Wphi}
\end{eqnarray}

In this particular coordinate system we would have to be careful because both vectors $V_{(3)}^{\alpha}$ and $V_{(4)}^{\alpha}$ before normalizing would be zero at the coordinate value $\theta = \pi / 2$. As the purpose of this section is not to find suitable coordinate coverings but to show that both the Coulomb and the Monopole electromagnetic fields are indispensable components of the tetrad vectors that make up the Reissner-Nordstr\"{o}m geometry, we are not going to look for other coordinate coverings. It is important to remark that we are fully aware of the singularity associated to the Monopole geometry, the different potentials that can be defined in different regions with the overlapping gauge transformation, see references \cite{BF}$^{,}$\cite{MN}$^{,}$\cite{WY} and chapter V Bis in reference \cite{CBDW}. For the point we are trying to make, it is enough to consider one of the regions where we have a unique definition of the Monopole potential. We do not need in this manuscript to describe the transition of the Monopole potential between different regions through a gauge transformation. We focus on just one region with one potential and study the relationship between the uniqueness of the local planes that diagonalize the stress-energy tensor and its relationship to the Coulomb and Monopole fields as simultaneously indispensable in the construction of local tetrad skeletons. Obviously if there is a transition to another region where another Monopole potential is defined through a local gauge transformation, there will be a local spatial rotation LB2 of unit tetrad vectors (\ref{Z}-\ref{W}) in this region of transition. The two vectors (\ref{Z}-\ref{W}) will undergo a spatial rotation like (\ref{ZT}-\ref{WT}) inside the local plane two in this region of transition, precisely \cite{BF}$^{,}$\cite{MN}$^{,}$\cite{WY}$^{,}$\cite{CBDW}.

\section{Conclusions}
\label{conclusions}

We are going to focus the summary of this work on two issues. First we notice that the constant associated both to the Coulomb and the Monopole field $q$ is the same. It is the Reissner-Nordstr\"{o}m charge constant. It cannot be any other way, otherwise the tetrads would not reproduce the Reissner-Nordstr\"{o}m metric tensor. Because the extremal field and its dual both carry the same charge. Conversely, nothing is said about the nature of this charge. It could be electric, magnetic, a combination. Second, the new tetrads have as it was said before, two main structure components. The skeleton, like $\ast \xi^{\alpha\lambda}\: \ast \xi_{\rho\lambda}$ in the tetrad vector (\ref{W}) and the gauge vector like $\ast A^{\rho}$ in the same vector, just to show an example. The skeletons are invariant under local electromagnetic gauge transformations because the extremal field and the metric tensor are local gauge invariants. Therefore, in this sense they have a unique expression. The dual to the Coulomb extremal field is the Monopole extremal field. When we build the two unit vectors (\ref{U}-\ref{V}), the extremal field, that is the Coulomb extremal field and the Reissner-Nordstr\"{o}m metric tensor are involved and necessary. When we build the two unit vectors (\ref{Z}-\ref{W}), the dual extremal field, that is the Monopole extremal field and the Reissner-Nordstr\"{o}m metric tensor are involved and necessary. The four orthonormal vectors make up a tetrad the locally and covariantly  diagonalizes the stress-energy tensor. They uniquely define blade one and blade two at every point in the region under study. The tetrad itself is not unique because we have the gauge freedom to choose the gauge vectors as $X^{\mu} = A^{\mu}$ and $Y^{\mu} = \ast A^{\mu}$, or add any local gauge transformation like $X^{\mu} = A^{\mu}+\Lambda^{,\mu}$ and $Y^{\mu} = \ast A^{\mu}+\ast \Lambda^{,\mu}$. The effect of this operation is on blade one either a local boost or a discrete transformation of the tetrad vectors (\ref{U}-\ref{V}), on blade two just a local spatial rotation of the tetrad vectors (\ref{Z}-\ref{W}), see reference \cite{A}. The point we are trying to make is that since the skeletons are unique, because there is a unique choice of skeletons that locally and covariantly diagonalizes the stress-energy tensor, then the Coulomb and the Monopole extremal fields must be components of the same geometry. Otherwise, there would be freedom in the choice of skeletons, and there is not, as long as we are trying to diagonalize the stress-energy tensor. In conclusion, the Reissner-Nordstr\"{o}m geometry is a unification structure within which the Coulomb and the Monopole coexist. Both are an integral part of this geometry. We quote from \cite{PD} ``The question arises as to whether an elementary particle can have both a charge and a pole \ldots It does not seem possible to answer the question reliably until a satisfactory treatment of the interaction of a particle with its own field is obtained.'' From the same manuscript \cite{PD} we also quote ``The field equations of electrodynamics are symmetrical between electric and magnetic forces. The symmetry between electricity and magnetism is, however, disturbed by the fact that a single electric charge may occur on a particle, while a single magnetic pole has not been observed to occur on a particle.''

%\bibliography{your-bib-file} % place the references here.

\end{document}